\documentclass[conf]{new-aiaa}
\usepackage[utf8]{inputenc}

\usepackage{graphicx}
\usepackage{amsmath}
\usepackage[version=4]{mhchem}
\usepackage{siunitx}
\usepackage{longtable,tabularx}
\usepackage{xspace} \usepackage{xcolor}
\newcommand	{\incfig}	[3]	{\begin{figure}[!t]
	\begin{center}
   	 \includegraphics
				[#2]	{fig-#1}
    	\caption{#3}
    	\label{fig:#1}
	\end{center}
	\end{figure}
	}

\newcommand {\citeref}
	[1] {\citep{RefNumber#1}}
	
\newcommand{\tc}[1]{\multicolumn{2}{c|}{$#1$}}

\newcommand{\secref}[1] {\S\ref{sec:#1}}
\newcommand{\tblref}[1] {Tbl.\ref{tbl:#1}}
\newcommand{\figref}[1] {Fig.\ref{fig:#1}}

\newcommand{\ile}[1]{\mbox{$#1$}}

\newcommand{\rate}{\mathcal R}

\newcommand{\SBR}{\text{SBR}}

\newcommand{\BPP}{\text{BPP}}

\newcommand{\blue}[1]{#1}
\newcommand{\levelone}{\blue{collector}\xspace}

\newcommand{\leveltwo}{\blue{aperture}\xspace}

\title{
Technological Challenges in Low-mass Interstellar Probe Communication
}

\author{David G. Messerschmitt}
\affil{Roger Strauch Emeritus Professor, Dept. of EECS,
University of California at Berkeley, USA}
\author{Philip Lubin}
\affil{Professor, Dept. of Physics, University of California at Santa Barbara, USA}
\author{Ian Morrison}
\affil{Research Fellow,
International Centre for Radio Astronomy Research, Curtin University, Australia}

\begin{document}

\maketitle

\begin{abstract}
Building on a preliminary paper design of a downlink from a swarm of low-mass interstellar 
probes for returning scientific data from the vicinity of Proxima Centauri, the most critical technology issues are summarized, and their significance is explained
in the context of the overall system design.
The primary goal is to identify major challenges or showstoppers if such a downlink 
were to be constructed using currently available off-the-shelf technology, and thereby provide direction and 
motivation to future research on the constituent design challenges and technologies. 
While there are not any fundamental physical limits that prevent such communication systems, 
currently available technologies fall significantly short in several areas and 
there are other major design challenges with uncertain solutions. 
The greatest identified challenges are in mass constraints, 
multiplexing simultaneous communication from multiple probes to the same target exoplanet, attitude control and pointing accuracy, 
and Doppler shifts due to uncertainty in probe velocity. 
The greatest technology challenges are electrical power, high power and wavelength-agile optical sources, 
very selective and wavelength-agile banks of optical bandpass filters, 
and single-photon detectors with extremely low dark-count rates. 
For a critical subset of these, we describe the nature of the difficulties we encounter and their origins in the overall system context.
A receiver that limits reception to a single
probe is also considered and compared to
the swarm case.
\end{abstract}

\section{Introduction}
\lettrine{A}{} near-term opportunity for exploration of nearby stars is low-mass probes accelerated to
relativistic speeds using a directed energy (DE) driven sail for propulsion
\cite{lubin1}, \citeref{867}.
Such a mission is worthwhile only if acquired scientific data can be returned to earth using a
communication downlink \citeref{1015}.
In a companion paper \citeref{833} we have performed a 
paper design and idealized theoretical evaluation of the feasibility of such a
data downlink.
Although we are not aware of any fundamental physical limits on the data rate at which scientific
data can be reliably recovered, serious
challenges arise when relating the requirements to resource limits (such as probe mass and
electrical power) and available technologies (such as sources, detectors, and filters).
Current technologies fall considerably short of minimum requirements in several areas, and
the feasibility of a downlink is dependent on enabling technology advances driven either
by the natural advancement in technology or specific technology developed in the
course of R\&D dedicated to this application.
These enabling technologies are needed
between now and the first launch (for transmitter)
or coinciding with the arrival of the first scientific data (for the receiver).
The primary goal of this evaluation was thus to identify the most critical technological
shortcomings that present an obstacle to a successful mission, and thus help to guide
future R\&D efforts addressing these shortcomings.
Here we summarize those critical technologies, emphasizing their role within the overall downlink system
based on the results from \citeref{833}.
This paper is based on \cite{messer3}, updating and adding some material.

\section{Nominal design}

A simplified block schematic of the optical signal path is shown in \figref{SignalPath}
as a basis for later discussion.
It illustrates the primary elements of that signal path discussed in this paper and how they relate to one another.
A model of the downlink has a number of configuration parameters and performance metrics.
A \emph{nominal} design invokes compatible (and hopefully reasonable) choices for the parameters listed in \tblref{params}.
A model of the downlink, including both physical layer (having to do with physical elements and propagation)
 and transport layer (having to do with modulation and coding), yields a set of
performance metrics listed in \tblref{metrics}.
The model is idealized, ignoring many practicalities such as pointing inaccuracy, aperture sidelobes,
and coding shortfalls, and thus yields optimistic values for all the metrics.\footnote{
See \citeref{833} for
plots of changes in metrics as parameters are varied one at a time.
Software code allowing the reader to substitute their own model parameters and calculate the resulting metrics is available at
\citeref{1021}.
}

\incfig
	{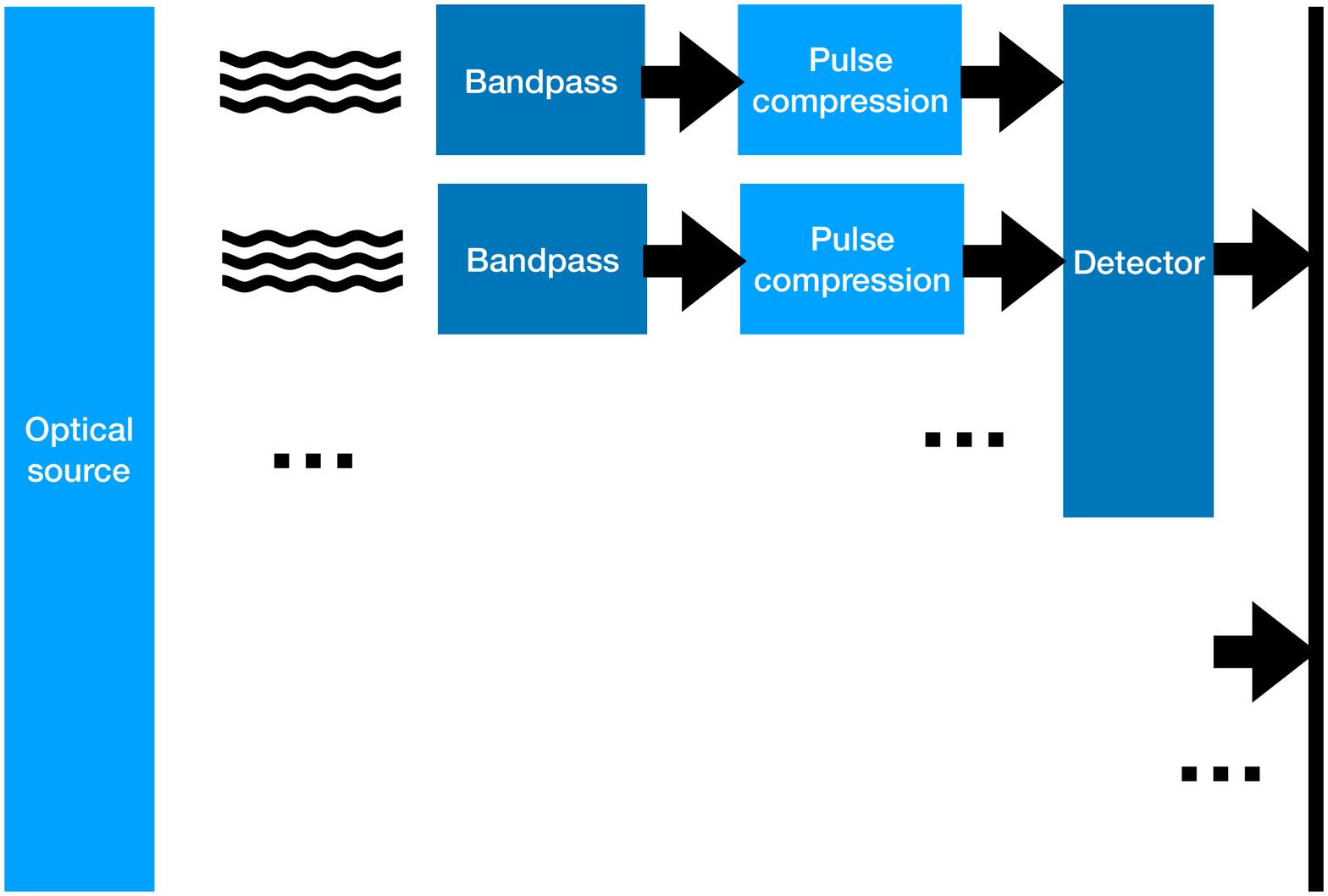}
	{
	trim=0 70 0 50,
    	clip,
    	width=.5\linewidth
	}
	{
	The important elements in the optical signal path from electrical power input to photon detection events at the output.
	The optical source generates pulses of optical power which propagate to the receiver,
	where a bandpass filter limits noise and interference and (possibly) separates signals from different probes.
	The pulse compression is a phase-only filter that concentrates signal pulses in time, moderating the
	peak power required at the optical source.
	The optical detector labels individual photon detection events with time stamps.
	Each detector may be shared over multiple receive \leveltwo{s} and will be replicated many times
	for different sets of \leveltwo{s}.
	}

 \begin{table}
 \caption{\label{tbl:params} 
 Design parameters and values for a nominal design
 including reception from a single probe and simultaneous reception from a swarm of probes.}
 \centering
 \begin{tabular}{|c |p{8cm} |c|c|c|}
 \hline
 Parameter & Description & Units & Swarm & Single
 \\ \hline
 $\rate_0$ 
 & Data rate, before outages, immediately following encounter
 & bits/s
 & 
\tc{1.}
 \\
 $D_0$ & Distance to target star (Proxima Centauri)
 & ly & 
\tc{4.24}
 \\
 $u_0$ & Speed of probe as fraction of speed of light $c$ && 
\tc{0.2}
 \\
$ \lambda_0$ & Receive wavelength & nm & 
\tc{400}
\\
$A_e^T$ & Effective area of transmit aperture & cm\textsuperscript{2} & 
\tc{100.}
\\ \cline{4-5}
$\Omega_A$ & Solid angle of receive \levelone coverage
& arcsec\textsuperscript{2} 
& 10. & 0.01
\\ \cline{4-5}
SBR & Signal-to-background-radiation ratio at each \leveltwo and for the receive \levelone as a whole
&& 
\tc{4.0}
\\
$\Lambda_D^S$ & Average rate of dark counts referenced to each \leveltwo
& ph/yr & 
\tc{32.}
\\
$\delta$ & Duty cycle for burst-mode transmission
&& 
\tc{1.9{\times}10^{-4}}
\\
$T_s$ & Individual pulse width at optical detector & ns & 
\tc{100.}
\\
$K_s$ & Average number of photon detections in each pulse && 
\tc{0.2}
\\
$W_e$ & Effective optical bandwidth (\ile{W_e T_s = 1}) & MHz & 
\tc{10.}
\\
$P_0$ & Outage probability (daylight and clouds)
&& 
\tc{0.57}
\\ \hline
 \end{tabular}
 \end{table}

The probe mass is assumed to be $\gamma$ grams, where \ile{\gamma{\ge}1} is the \emph{mass ratio}.
For a fixed launch infrastructure, increasing $\gamma$, while reducing $u_0$, also increases $\rate_0$
due to larger transmit aperture and electrical power.
An optimum choice of $\gamma$ minimizes the total \emph{data latency} (time elapsed from
launch to completion of scientific data download) for a given \emph{data volume} 
(total number of bits or bytes downloaded and recovered reliably).
For a particular set of scaling laws, we find that
transmission should be terminated after the probe reaches a distance of
approximately $1.10{\cdot}D_0$, with a resulting nominal transmission time of 2.12 yr.

The signal-to-background ratio SBR is the ratio of average signal photon rate to average background radiation
photon rate for all sources of background including unresolved sources of noise
(Zodiacal radiation, deep star field, and scattered moonlight), a point source of interference radiation from the
nearby target star, dark counts in the received optics and the optical detector, 
and incomplete extinguishment of the optical source
during ``off'' periods.
With the notable exception of dark counts, the noise and interference
power is limited by a narrow optical bandpass filter with bandwidth \ile{W_e{=}10} MHz before it contributes to SBR.
We anticipate using superconducting optical detectors 
shared over multiple \leveltwo{s} in order
to achieve the very low dark count rate $\Lambda_D^S$
(see \secref{detectors}).
The chosen objective for SBR is sufficiently large that theoretical limits indicate minimal impact on the data rate at which
reliable detection can be achieved.
During nighttime we found that scattered moonlight and dark counts are the most significant sources of background radiation for typical interference parameters.

Theoretical limits also indicate that reliable data recovery can be achieved by constructing a signal at the
detector composed of narrow pulses of duration $T_s$, with the average number of photons detected $K_s$ within each
pulse.
The small value \ile{K_s{=}0.2} (termed photon starvation mode)
 is advantageous in minimizing peak transmit power,
but it indicates that a random ${\sim}80$\% of the pulses are unobservable by the receiver because zero photons are detected.
Reliable extraction of scientific data is achieved
by error-correction coding (ECC), which adds
substantial redundancy to the scientific data before transmission.
In the nominal design, 17\% of the total transmitted data is scientific and 83\% is redundant information used for error control
and to counter random outages.
This design approach has been validated in bench testing \citeref{834}.

Scattered sunlight during the daytime is far too large to overcome, and thus daylight 
as well as adverse weather events are treated as a receiver-initiated outage.
The total outage probability $P_0$ in \tblref{params} assumes
a daylight outage probability (averaged over a year) of 52\% and a weather outage probability of 10\%.
Through a combination of interleaving and ECC,
a post-outage data rate approaching \ile{\rate_a{=}\rate_0{\cdot}(1 - P_0 )} 
(nominally 0.43 bits/s with reliable recovery of scientific data) can be achieved,
and significantly the probe need not have knowledge of outage conditions.

Dark counts are perverse photon-detection events unrelated to incident signal, noise, and interference,
and may occur in the optics and photonics (due to blackbody radiation) as well as optical detectors
\citeref{1010}.
Although it is clear that superconducting optical detectors are required, and they fortunately have no intrinsic
mechanism for dark counts,
nevertheless occasional detector dark counts will be caused by non-idealities such as impurities and radiation.
Since the achievable dark count rate $\Lambda_D$ is largely technology driven,
there is no solid rationale for choosing a specific numerical value.
The nominal value in \tblref{params} is chosen
to render the dark-count contribution to background radiation equal to that of scattered moonlight (for a full moon)
within the context of the other design parameters.
Moonlight and dark counts are then the two dominant sources of background radiation in equal measure.

\section{Receiver issues}

The technology issues that arise in the receiver can only be addressed in the context of the
receive \levelone architecture, so that is briefly described 
before addressing the optical processing and detector technology challenges.

\subsection{Coverage}

Communications is much easier for a single probe heading to a single target but we also address the more complex issue of simultaneously receiving communications from a swarm of probes all headed to the same target but whose launches are separated in time enough to lead to an angular spread in probes requiring a larger instantaneous field of view or \emph{coverage}.
A larger coverage results in a lower receiver sensitivity \citeref{822}.
Two limiting cases are listed in \tblref{params}:
\begin{itemize}
\item
For the ``swarm'' case,
the relatively large coverage solid angle $\Omega_A$ in \tblref{params} allows a single receive \levelone
to capture a superposition of signals from all the probes in a swarm that are transmitting simultaneously.
As illustrated in \figref{parallaxTraj}, parallax and proper motion of the star cause this coverage to expand
as the number of probes operating downlinks concurrently increases.
For this reason, downlink operation for other than scientific data purposes should be kept to a minimum,
and the duration of downlink operation (and hence total data volume) should also be minimized.
The optimum downlink operation time is found to be about 10\% of the transit time, which is 2.12 yr.
With a 30-day inter-launch interval, there are 26 downlinks operating at any given time.
\item
For the ``single'' case, coverage is assumed to be limited to a single probe.
That coverage will be lower bounded by the pointing accuracy that can be achieved.
With adaptive optics, it is assumed that a pointing accuracy of approximately 0.1 arcsec may be
achievable, with the conclusion that a coverage solid angle of about 0.01 as\textsuperscript{2}
will ensure that the probe trajectory falls within the coverage area.
\end{itemize}
There are a number of scenarios that fall between these limiting cases.
For example, the coverage could be
reduced with a more rapid launch sequence interspersed with a launch holiday
larger than the transmission time, thereby limiting the effect of star proper motion.

\incfig
	{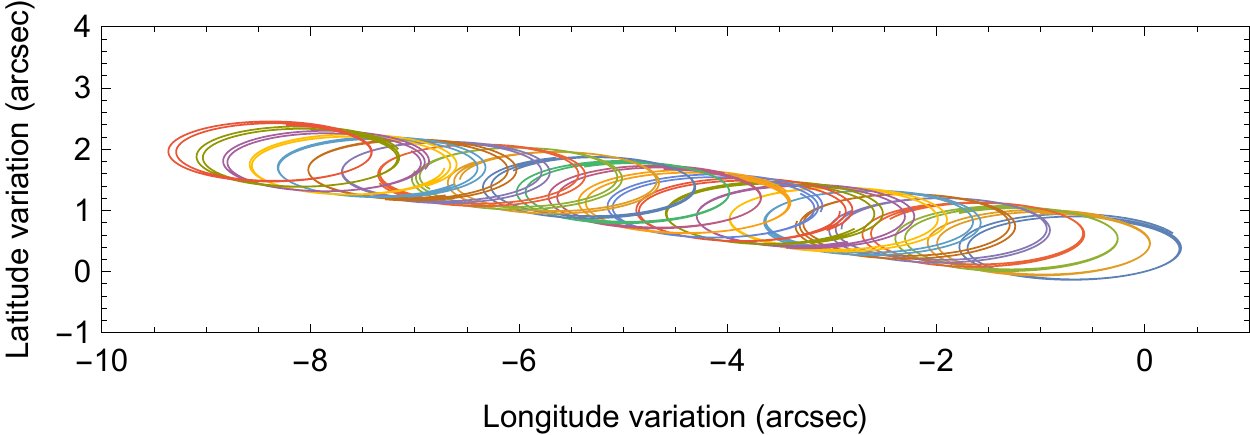}
	{
	trim=0 0 0 0,
    	clip,
    	width=.8\linewidth
    	}
	{
	Relative angle of probe trajectories as seen from a terrestrial receiver.
	Shown in different colors are the trajectories over 2.12 years of downlink operation for each of 26 probes
	launched at 30 day intervals.
	The oval shape for each probe's trajectory is due to the parallax effect as the probe as viewed from different locations on the
	earth's orbit.
	The general drift in the trajectories is due to the proper motion of
	the target star Proxima Centauri, which requires the launch angle of the probes to change 
	so as to track the target.
	}

\subsection{Receive \levelone decomposition}
\label{sec:decomposition}

The terrestrial receiver for the downlink captures the receive optical power
using an optical \emph{\levelone}.
A km-scale receive \levelone is required to achieve sufficient detected photons
to allow reliable extraction of the scientific data.
A single-mode diffraction-limited aperture of this size would not only be impractically
large to implement, but the pointing accuracy requirement would also be impractical
and the aperture would not be able to receive data from more than a single probe at a time.
These issues are all overcome by decomposing the entire \levelone into replicated
\leveltwo{s} as illustrated in \figref{ReceivedApertureSimplified}.
To clearly distinguish parameters and metrics, variables are labled with a ``T'', ``R'', or ``S'' respectively depending on whether
they apply to the transmitter, receive \levelone in its entirety, or are referenced to each \leveltwo.
Thus, in \tblref{metrics} the effective area of each \leveltwo, 
assumed to be diffraction-limited, is $A_e^S$, and the number of \leveltwo{s} is $N^S$.
The \leveltwo parameters govern the sky coverage $\Omega_A$ of the receiver, and also influence the SBR.
A fundamental antenna theorem tells us that for uniform coverage over a solid angle $\Omega_A$ (ideally with complete rejection
of all other angles) and effective area $A_e^S$, we must have \ile{A_e^S \cdot \Omega_A = \lambda_0^2} \citeref{822}.
$A_e^S$ is an area measure capturing the \leveltwo sensitivity (ratio of detected power to incident 
plane-wave electromagnetic flux),
and the sensitivity and coverage are thus inversely related.

\incfig
	{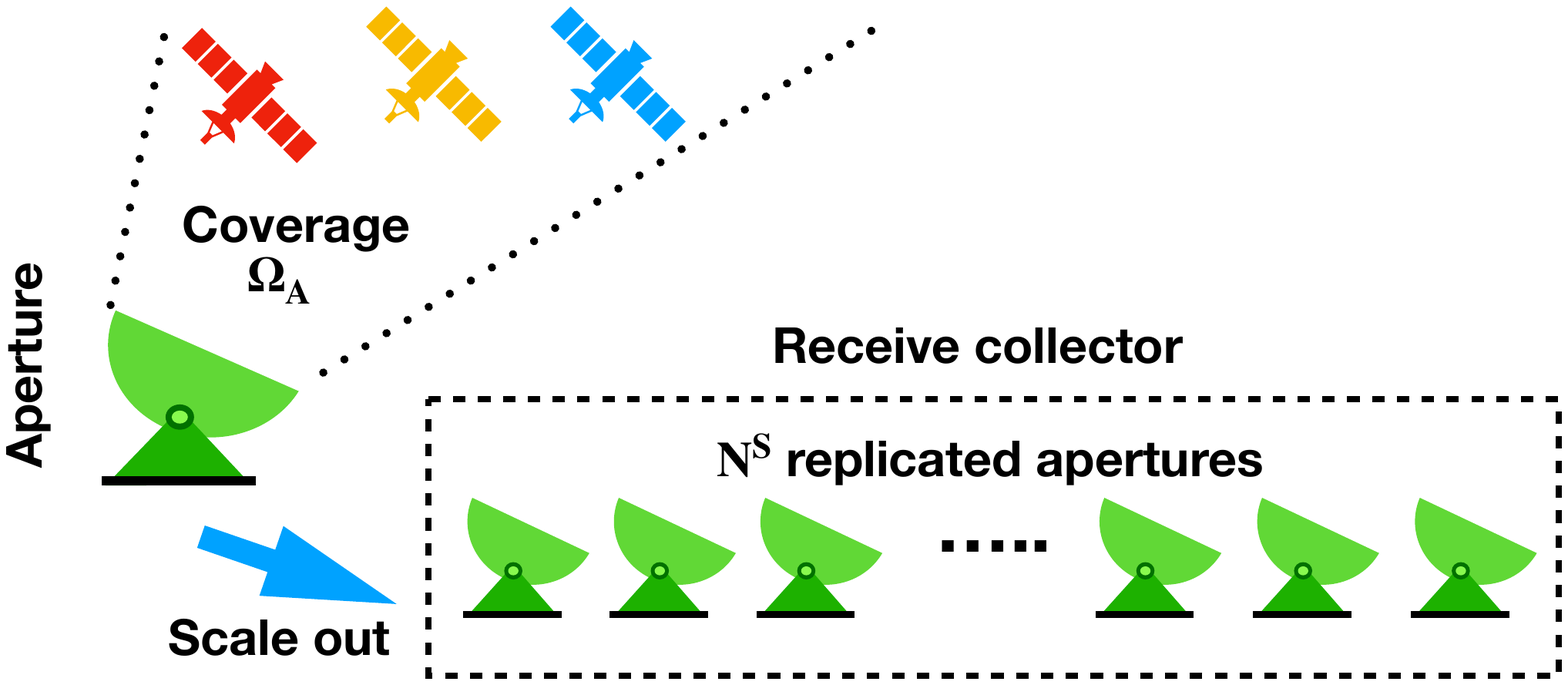}
	{
	trim=0 330 0 0,
    	clip,
    	width=.8\linewidth
	}
	{
	The large receive \levelone is decomposed into a large number $N^S$ of \leveltwo{s}.
	Each \leveltwo controls the coverage solid angle $\Omega_A$, and the average transmit power $P_A^T$ is
	chosen to achieve the desired signal-to-background ratio SBR at the output of each \leveltwo.
	The \leveltwo{s} operate independently and add signal photon detection events incoherently so that
	neither $\Omega_A$ nor SBR is affected.
	}

\begin{table}
 \caption{\label{tbl:metrics} 
 Performance metrics for the parameters of \tblref{params}. 
 Average transmit power scales in proportion to data rate while
 other metrics remain unchanged.
 }
 \centering
 \begin{tabular}{|c |p{8cm} |c |c |c|}
 \hline
 Parameter & Description & Units & Swarm & Single
 \\ \hline
 BPP & Photon efficiency with reliable recovery of scientific data
& bits/ph & 
\tc{10.9}
\\
$\rate_a$ & Actual scientific data rate following outages
& bits/s & 
\tc{0.43}
\\ \cline{4-5}
 $A_e^S$ 
 & Receiver: Effective area of an \leveltwo
 & cm\textsuperscript{2}
 &6.8 & 6800.
 \\
 $N^S$ & Receiver: Number of replicated {\leveltwo}s
 &  & $5.9{\times}10^7$ & $2.2{\times}10^6$
 \\
 $N^S{\cdot}A_e^S$ & Receiver: Total of \leveltwo effective areas & km\textsuperscript{2} & 0.04 & 1.5
\\
$P_A^T$ & Transmitter: Average transmitted power & mW & 29.4 & 0.8
\\ \hline
 \end{tabular}
 \end{table}

$A_e^S$ is a metric determined by the choice of parameter $\Omega_A$.
As a result each \leveltwo is sub-meter-scale, and hence achieving close to diffraction-limited performance is feasible.
The photon events from the multiple \leveltwo{s} are added incoherently, and thus the replication of $N^S$
\leveltwo{s} is a budgetary and operational (but not technological) issue.\footnote{
All that \leveltwo{s} have in common is a clock for time stamping of photon events.
The relative (not absolute) clock tolerance has to be about 1 ns for the nominal design,
which uses pulses with 100 ns duration at the optical detector input.
}
Since both signal and background components add incoherently in proportion to $N^S$, the
SBR as well as coverage are transparent to \leveltwo replication.

The background radiation denominator of SBR at each \leveltwo is governed by
the radiation intensity, effective bandwidth $W_e$ (in the case of noise and interference, but not dark counts),
and \leveltwo area $A_e^S$ (only in the case of interference\footnote{
The output power of each \leveltwo for unresolved sources of noise is proportional to $A_e^S$
(the effective collection area) and $\Omega_A$ (the sky solid angle over which which noise enters the \levelone).
By the antenna theorem the product of these two factors is a constant, and hence noise output power
is not affected by coverage $\Omega_A$ \citeref{822}.
}).
Each \leveltwo signal photon-count output is proportional to $P_A^T A_e^T / \lambda_0$, and
minimizing $P_A^T$ provides one motivation for choosing a short optical wavelength
(at radio wavelengths, for example, either the transmit power or the receive \levelone would have to be impractically large).
A second motivation is minimization of the interfering background radiation from Proxima Centauri (a red dwarf star),
the spectrum of which is decreasing rapidly with shorter wavelengths in this region.

In the ``swarm of probes'' case in \tblref{metrics},
the nominal number $N^S$ of \leveltwo{s} is large (59 million) and their total effective area adds up to 4\% of a km\textsuperscript{2}.
It is likely that this large number of apertures would be composed of sub-assemblies, each
such sub-assembly containing an array of apertures manufactured as a unit. 
In the ``single probe'' case, assuming an achievable 0.1 arcsec pointing accuracy, the effective area $A_e^S$ of each
\leveltwo is larger and the total effective area is 1.5  km\textsuperscript{2}.
In return for this tighter coverage and commensurately larger area, the average transmitted power can be considerably smaller
(0.8 vs 29.4 mW) or for the same average power the data rate for the single probe can be correspondingly higher.
There is thus a price to be paid for simultaneous coverage of a swarm of probes in the electrical power requirement at each probe,
but a compensating payoff in a smaller total terrestrial \levelone area.

\subsection{Design partitioning}

The architecture of \figref{ReceivedApertureSimplified} is dictated by the principles of antenna theory, 
and admits a conceptually simple
functional partitioning of the design:
\begin{itemize}
\item
The data rate $\rate_0$ is determined by the rate of bursts in transmission (see \secref{burst}).
\item
The reliability with which data is recovered at the receiver is determined by the energy associated with each burst.
Since this energy requirement remains fixed, the transmitter  power-area product $P_A^T A_e^T$ is proportional to  $\rate_0$.
\item
The coverage solid angle $\Omega_A$ is determined by the effective area $A_e^S$ 
of each \leveltwo in a \levelone,
which in turn determines the receiver sensitivity at each \leveltwo.
Thus the sensitivity of the \leveltwo is determined by the \levelone coverage.
\item
For a fixed $\rate_0$,
the SBR at each \leveltwo necessary to achieve reliability in data recovery determines the minimum
transmitter power-area product $P_A^T A_e^T$.
\item
Achieving a sufficient photon count $K_s$ in each burst for reliable data recovery at the entire receive \levelone output 
determines the minimum the number $N^S$ of \leveltwo{s} and thus the total receive \levelone effective area $N^S A_e^S$.
\item
It is possible to trade off total receive \levelone effective area $N^S A_e^S$ for lower peak and average power at the probe.
This is accomplished indirectly by reducing the photon efficiency BPP.
(The primary effect of reducing BPP is to increase the receive \levelone area, but it also reduces the required transmitted power through a reduction in the background radiation
due to a smaller optical bandwidth.)
\end{itemize}
Two unavoidable and coupled constraints, the determination of \leveltwo effective area by coverage and the accumulation of
dark counts with multiple \leveltwo{s}, strongly influence the architecture and design.
Together they eliminate some expected trade-offs such as 
trading a higher transmitted average power for a reduced transmit peak power (see \secref{peakPower}).

\subsection{Optical detector and dark counts}
\label{sec:detectors}

Only very infrequent dark counts can be tolerated as an unintended side effect of the low 
data rate $\rate_0$
and high photon efficiency BPP.
We can better understand the nature of this challenge by examining the signal photon
rate for the nominal values in Tbls.\ref{tbl:params} and \ref{tbl:metrics}.
The average photon detection rate overall is \ile{\rate_0/\BPP{=}0.09} ph/s,
and to achieve the SBR objective (assuming dark counts are the only source of background radiation) we must have
\[
\SBR =
\frac
    {\rate_0 / \BPP}
    {N^S \cdot \Lambda_D^S} 
\ \ \ \text{or}\ \ \ 
\Lambda_D^S = 1.2\ \text{ph/century} \,.
\]
where $\Lambda_D^S$ is the dark count rate
referenced to one \leveltwo.
This clearly cannot be achieved in a
straightforward manner.
The main challenge comes from the large $N^S$, which results in an accumulation of dark counts
when each \leveltwo has an independent optical detector.
Clearly we must find ways to mitigate a higher dark count rate.

\subsubsection{Burst mode transmission}
\label{sec:burst}

A first dark count mitigation strategy is to transmit in \emph{burst mode}.
The idea for this follows from the observation that  we are not only challenged by
a large $N^S$, but also by the small $\rate_0 / \BPP$.
Most sources of background radiation benefit from the lower bandwidth
associated with a low data rate.
This is also true of dark counts due to blackbody radiation in the optics and photonics,
but dark counts originating in the optical detector are a notable exception.
To help overcome this, as shown in \figref{BurstMode} we can artificially increase the instantaneous
data transmission and signal photon rate by limiting
transmission to short bursts interspersed with blanking intervals.
During blanking intervals the receiver ignores all photon counts, including all sources of background radiation.
This effectively reduces the dark count rate by the duty cycle $\delta$, which is
the fraction of time occupied by the data bursts.
The nominal design achieves a \ile{\delta^{-1}{\sim}5300} reduction in dark counts.
Background photon detection events due to noise and interference are also
reduced by factor $\delta$, but in this case there is no net effect because
an increase in optical bandwidth \ile{W_e \sim \delta^{-1}} admits more photons during the data bursts.
This increase in optical bandwidth is beneficial in its own right since it relaxes the required selectivity of the optical bandpass filters.

\incfig
	{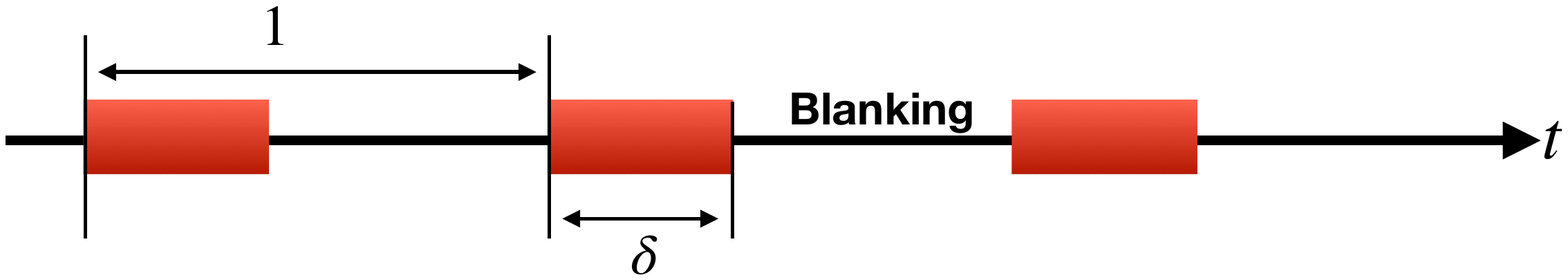}
	{
	trim=0 470 0 0,
    	clip,
    	width=.8\linewidth
	}
	{
	An illustration of burst mode transmission.
	Data is transmitted in bursts with a duty cycle $\delta$ interspersed
	with blanking intervals during which the receiver ignores all photon events.
	The effective data rate during bursts is increased by factor $\delta^{-1}$
	and the dark count rate is effectively multiplied by $\delta$.
	}

\subsubsection{Multiple \leveltwo{s} per optical detector}

Even after benefiting from burst mode, the dark count rate in \tblref{params} 
may be beyond the capability of any
current superconducting optical detectors if we maintain a one-to-one correspondence between \leveltwo{s} and detectors \citeref{1010}.
Technological progress in detectors should be feasible, including enhancing the purity of materials, radiation shielding,
and techniques for eliminating some spurious photon detections.
A second dark count mitigation strategy would be to share each optical detector over multiple \leveltwo{s}
as pictured in \figref{SignalPath}.
The resulting reduction in the total number of optical detectors would reduce the total dark count rate.
However, the incoherence of optical signals across \leveltwo{s} must be maintained
to ensure that the coverage and SBR are not modified by this sharing.
Thus optical interference among \leveltwo optical outputs prior to detection must be avoided.

\subsection{Optical processing}

Optical processing is the manipulation of the optical signal prior to detection.
To avoid optical interference between \leveltwo{s}, it is likely that this processing would
need to be dedicated to each \leveltwo (as opposed to each optical detector).
One important function is optical bandpass filtering to limit the noise and interference
reaching the detector.
In the nominal design, the use of burst mode increases the optical bandwidth of the
signal from \ile{W_e{=}1.9} kHz to 10 MHz, beneficially reducing the selectivity requirement
for the filter.
The resulting frequency selectivity is one part in $7.5{\times}10^7$, which is within reach
\citeref{879}.

A major complication to the optical processing is the superposition of
received signals due to multiple probes in a swarm of probes transmitting concurrently,
which is called \emph{multiplexing}.\footnote{
Even when a receive \levelone is dedicated to each probe there will be an additional source of interference
from other probes which occupy an overlapping bandwidth.
}
Multiplexing places the additional burden
on the optical processing and/or modulation coding to separate the signals from different probes
before the remainder of the receiver processing.
There are at least four different multiplexing techniques based on separation of signals
by angle, by frequency, by time, or by code.
We have not made a specific multiplexing proposal. This is a major issue affecting
not only optical processing but many other aspects of the design.
On initial examination time-division multiplexing seems a relatively feasible option due again to
the low duty-cycle $\delta$ of burst mode.
Unfortunately, synchronization of bursts between different probes is not feasible,
as it would require an impractically precise clock on the probe.
However, an Aloha-type protocol such as employed in ethernet \citeref{890},
 in which the temporal character of
transmitted data bursts are deliberately randomized, may be
feasible, albeit at the expense of an additional source of outages due to the infrequent
random overlapping of bursts.

No matter what choices are made for \leveltwo-sharing and multiplexing, Doppler shift
will be a significant issue due to the inevitable uncertainty and tolerance in probe velocity.
Because of the relativistic speed in combination with high carrier frequency, this is a significant effect.
For the nominal parameterization in \tblref{params}, for each $\pm 1$\% uncertainty in 
probe speed there is a $\pm 1.6$ THz Doppler shift in receive frequency.
There is thus uncertainty in the receive wavelength from each probe,
and different probes will have different Doppler shifts.
This would be an issue to overcome in either time- or wavelength-division multiplexing.
There are a combination of countermeasures that can be taken, such as 
launch control for precise cruise velocity,
a telemetry uplink communications
that operates shortly after
launch for a compensating configuration of transmit wavelength, 
as well as possibly a need for agility and configurability
in the center frequency of the receive optical bandpass filters.

\section{Probe transmitter technologies}

The transmitter is constrained by the aggressive mass limitation necessary to
achieve a relativistic probe speed.
Thus the technological challenge is largely one of miniaturization of functions
that could be readily achieved on a terrestrial laboratory bench.
Where functionality is partitioned between transmitter and receiver, we obviously
want to place the greater burden on the receiver.
For example, the transmit aperture area $A_e^T$ is chosen to be sub-meter-scale in \tblref{params},
necessitating a km-scale receive \levelone in \tblref{metrics}.

\subsection{Electrical power supply and average power}

There are three non-overlapping phases of a mission where the available electrical power will be limiting:
Scientific data collection, data preparation prior to operation of the downlink (such as data
compression, addition of error-correction redundancy, and interleaving), and transmission back to earth.
In particular, the electrical power during transmission must be larger than the average transmit power $P_A^T$, with
a nominal value of 30 mW in \tblref{metrics}.
As this power is directly proportional to the scientific data rate $\rate_0$, it could be
reduced at the expense of reduced total data volume.

The power-area-area metric $P_A^T A_e^T \cdot N^S A_e^S$ is proportional to
$\BPP^{-1}$, and thus is moderated by achieving a high photon efficiency (BPP in bits/ph).
The transmission is based on pulses with too few detected photons $K_s$ to be recovered reliably at the receiver.
The design makes up for this by adding large amounts of redundancy, which allows scientific data to be recovered reliably.
It thus trades off energy savings in individual pulse transmission/reception for
the added power of transmitting the redundant data alongside the scientific data.
This approach may seem surprising, but is theoretically necessary to approach theoretical limits on BPP \citeref{798}.

Achieving a high reliability in scientific data recovery is the biggest challenge, with a typical requirement of one error on average
in 1-10 megabits.
That is the crucial role of ECC.
If reliable data recovery is not demanded, there is no theoretical limit on the $\rate_0$ that could be achieved.\footnote{
There is some confusion on this point in the physics literature, where occasionally it is incorrectly represented
that there is a theoretical limit on the data rate that can be achieved without consideration of the reliability of data recovery.
In fact, absent a reliability requirement any data rate can be achieved, as illustrated in the limit 
of poor reliability by simply
recovering data by a sequence of "random coin tosses" at the receiver. 
}

\subsection{Peak transmitted power}
\label{sec:peakPower}

The nominal design is based on pulses that are narrow (\ile{T_s{=}100\text{ ns}})
and sufficiently energetic to result in a meaningful number of photon detections (\ile{K_s{=}0.2}) across
the entire receive \levelone (all \leveltwo{s}).
A crucial point is that these narrow energetic pulses are required at the receiver optical detector inputs, but 
commensurately narrow pulses are \emph{not} necessarily required at the transmitter optical source.
If these narrow pulses were generated in straightforward fashion by an optical source in the
transmitter, the nominal peak transmit power would
be large (\ile{P_P^T{=}638\text{ kW}} for swarm reception and \ile{P_P^T{=}17.4\text{ kW}} for single-probe reception).
These values are likely not feasible with a miniaturized semiconductor laser, and further
converting a continuous-electric power source into narrow pulses would be an issue given the mass limitation.

Before describing our proposal for achieving the required pulses at the detector, let's address why
such energetic pulses are theoretically required
at the optical detector for the purpose of achieving a large photon efficiency BPP.
Avoiding a long theoretical development, this can be summarized intuitively in the following manner.

The number of individual photon detections is random, as dictated by quantum mechanics.
At the design goal of \ile{\BPP{=}10.9\text{ bits/ph}}, each scientific data bit recovery is
based on an average of 0.092 photon detection events.
Given the quantum-mechanical randomness in photon detections,
such a small number of photon detections is counter to the goal of reliable data recovery.
The only way to achieve reliability is to recover multiple bits based on a commensurate large number of photon detections.
Thus \ile{\BPP{=}10.9} might actually be achieved by exchanging an average of 100 detected photons for 1090 bits, leaving room for
the law of large numbers to dramatically improve the reliability in this exchange.
This simultaneous recovery of 1090 bits is another way to describe the methodology of ECC.
With this in mind, there are three intuitive explanations for the role of narrow pulses
at the photon-counting detector:
\begin{itemize}
\item
The most photon-conserving way to represent data is by the timing (as opposed to amplitude) of pulses,
and the narrower the pulses the more data
that can be represented by a single pulse in this manner.
\item
If we represent 1090 scientific data bits by 100 photon detection events on average as in our conceptual example,
at a data rate of \ile{\rate{=}1\text{ bits/s}} this implies 100 photon detection events within 1090 sec, or 18 minutes.
More significantly it implies that there must be  \ile{2^{1090}({\sim}10^{328}}) distinguishable patterns of pulses,
one pattern for each possible combination of 1090 bits.
Compare this to the \ile{{\sim}10^{80}} atoms in the visible universe.
Fitting that huge number of distinctive patterns into 18 min is only possible if the pulses are themselves narrow to start with.
\item
All else equal, reducing the transmission to bursts interspersed with blanking intervals,
though eliminating almost all dark counts, also reduces the pulse widths by $\delta^{-1}$.
This offers a direct tradeoff between dark count rates and pulse width that can be exploited depending on which technology
achieves the greatest advances in the future.
\end{itemize}
The overall effect of data reliability and dark count mitigation on pulse widths can be seen in the 
nominal bandwidth \ile{W_e {=}10} MHz, which is
seven orders of magnitude greater than the data rate.
This bandwidth expansion is attributable to the use of sparse 100 ns pulses in a transmission
with low duty-cycle.

\subsubsection{Design parameter adjustment}

The large peak powers that would be required at the transmitter to generate relatively narrow 100 ns pulses
illustrate the technical difficulty and tradeoff between peak transmitted power and receiver dark count rates.
Adjusting other parameters of the design can also increase pulse width $T_s$, thus easing the burden
on the transmitter optical source.
In particular,
burst mode transmission offers a direct tradeoff between $\Lambda_D^S$ and $T_s$,
and hence peak power generation at the transmitter.
Each order of magnitude reduction in $\Lambda_D^S$ and compensatory order of magnitude increase in $\delta$ 
permits $T_s$ to increase by an order of magnitude, all else equal.
Reducing the coverage $\Omega_A$ will also reduce the required pulse energy,
although this comes at the expense of a larger total receive \levelone area
and the need to replicate receivers, requiring in the limit a separate receiver for each probe in a swarm.

Reducing the photon efficiency objective must be compensated by a larger total receive \levelone
effective area $N^S A_e^S$.
One beneficial side effect of exercising this tradeoff turns out to be a reduction in transmitted
average and peak power due to a resulting decrease in total background radiation
due to a decrease in optical bandwidth.
Thus reducing photon efficiency in the context of burst mode and the \levelone architecture of
\figref{ReceivedApertureSimplified} implicitly
exchanges larger total \levelone area for smaller transmit average and peak power.

Our approach in \tblref{params} is to choose a signal level larger than the
background level (\ile{\SBR{>}1}).
It is also feasible to operate with \ile{\SBR{<}1}, although the theoretical photon efficiency
falls off rapidly as SBR decreases in this regime.
While this would allow for greater dark count rates, all else equal, it would have other
negative consequences.

\subsubsection{Employing optical interference}

Optical interference can be used in place of emitting high powers directly from an optical source.
A straightforward example of this is the ganging of multiple semiconductor lasers in the transmitter with an optical combiner.
Unfortunately this is expected to be inconsistent with the low-mass objective.

Pulse compression technology, often used to generate narrow laser pulses, converts a longer-duration waveform,
albeit a waveform with the same bandwidth as the narrow pulses, into narrow pulses by utilizing wavelength-dependent
group delays to align shorter- and longer-frequency components in time.
This would have to be done in the optical realm, and could presumably be accomplished in either transmitter
or (more interestingly) receiver.
Typical arrangements utilize pairs of diffraction gratings or prisms to provide the wavelength-dependent variable group delays.
While this technique is often used to generate picosec pulses, it does not appear to be feasible for the microsec-scale
pulses in our application due to the microsec-scale delays that would be required and the resulting large physical scale
(light travels a meter in ${\sim}3$ nanosec in a vacuum).
Burst mode offers the freedom to reduce pulse widths arbitrarily, bringing us into the feasible
range of pulse compression, but at the expense of a commensurately larger peak-power
requirement.

\subsubsection{Changing the modulation format}

The model used to quantify the metrics in \tblref{metrics} 
assumes a modulation code which operates at a single wavelength and represents scientific
data by the timing of narrow pulses in transmit and received power.
This is only one of a class of theoretically equivalent modulation codes based on
orthogonal codewords in the space of temporal and wavelength modes.

A second approach is
 to represent data by different wavelengths rather than different times.
 While this requires a wavelength-agile optical source at the transmitter,
a significant advantage is a constant transmit power.
However, the complexity of this frequency-shift keying receiver is significantly increased, as
a high spectral resolution would be required
with optical detectors replicated at the distinctive wavelengths.
Such a proliferation of detectors would appear to multiply the dark count rates accordingly.

A third approach to modulation is code-division keying, in which a set of orthogonal codes 
(rather than times or wavelengths) are used \citeref{889}.
These codes can also be designed with constant power over the codeword, and
they are distinguished at the receiver by filters matched to each codeword utilizing again optical interference.
Again an entire bank of optical detectors is required, one detector for each matched filter,
and there is therefore the expectation that dark counts are multiplied.

\section{Conclusions}

There are a considerable number of obstacles to achieving the downlink objectives with a focus on a large multiple probe swarm.
We have outlined the most troublesome ones identified to date, suggesting considerable need and opportunity for R\&D efforts
directed at overcoming these obstacles.
Readers with relevant expertise are encouraged to tackle these challenges.
While there is some opportunity for architectural and system innovations, the major need 
is early feasibility demonstration of novel
photonics technology and architectures for high-peak power optical sources, 
very frequency-selective optical filters (including delay equalization and band limitation) and
optical detectors with very low dark count rates.
Also required is an early and credible proposal for, and choice of, a modulation coding and multi-probe multiplexing strategy.

\section*{Acknowledgments}

PML gratefully acknowledges funding from NASA NIAC
NNX15AL91G and NASA NIAC NNX16AL32G for the NASA Starlight program
and the
NASA California Space Grant NASA NNX10AT93H,
a generous gift from the Emmett and Gladys
W. Technology Fund, as well as support
from the Breakthrough Foundation for its StarShot
program.
More details on the NASA Starlight program can be found
at \url{www.deepspace.ucsb.edu/Starlight}.

\end{document}